\documentstyle[aps,preprint,prl]{revtex}
\begin{document}                
\draft
\preprint{ }
\title{Breakdown of Landau Fermi liquid properties in the $2D$
BOSON-FERMION model}
\author{J. Ranninger and J. M. Robin }
\address{Centre de Recherches sur les Tr\`es Basses Temp\'eratures,
Laboratoire Associ\'e \`a l'universit\'e Joseph Fourier, Centre
National de la Recherche Scientifique, BP 166, 38042 Grenoble
C\'edex 9, France}
\maketitle

\centerline{ ( \today )}

\begin{abstract}
We study the normal state spectral properties of the fermionic
excitations in the
Boson-Fermion model.
The fermionic single particle excitations show a flattening of
the dispersion as the Fermi vector ${\bf k}_{_F}$ is approached from below,
forshadowing a Bogoliubov spectrum of a superconducting ground state.
The width of the quasiparticle excitations near ${\bf k}_{_F}$ increases
monotonically as the temperature is lowered.
In the fermionic distribution function this temperature dependence
is manifest in a strong modification of $n({\bf k})$ in a small
region below ${\bf k}_{_F}$, but a nearly $T$ independant
$n({\bf k}_{_F})$.
\end{abstract}

\vspace{3cm}

\pacs{Keywords: A. high $T_{_C}$ superconductors, D. electronic states,
                E. electron emmission spectroscopy}

\narrowtext

Intense theoretical work and controversy concerning the foundations for a
novel non-Fermi liquid metallic ground state has followed the
suggestion\cite{ref1} that in high $T_{_C}$ superconductors the
normal state might not behave as a standard Fermi liquid  (FL).
In attempting to fit the various anomalous thermodynamic and transport
properties of these materials, the concept of a marginal FL was
proposed\cite{ref2} in which the imaginary part of the fermionic
self-energy is supposed to vary like $| \omega - \epsilon_{_F} |$ rather
than like $( \omega - \epsilon_{_F})^{2}$ as in an ordinary FL.
Apart from this phenomenological approach to non-FL behavior, it is
well established that $1D$ systems  show in fact a branch cut
spectrum rather  than poles for the single particle Green's function.
For interacting Fermion systems such a non-FL behavior for $1D$ is
however not maintained if one goes to $D > 1$\cite{ref3}.

Since possibly the non-FL properties of the high $T_{_C}$ materials are
a prerequisite for the large values of $T_{_C}$ themselves\cite{ref4} and
also since the concept of a non-FL in itself is of fundamental
theoretical importance, it is of great interest to study models which
exhibit transitions between FL and non-FL behavior as some parameter of the
model is varied.

In anticipation of a breakdown of FL behavior for the Anderson impurities
lattice model with finite range interaction,
the single impurity problem
has been studied\cite{ref5}. When the repulsive interaction is increased
beyond a certain threshold
a change-over from FL to non-FL is indeed observed due to the coupling
to the ensemble of the
orbital channels rather than to only the one which hybridizes with the
impurity orbital.

There are so far few attempts to study models in which such effects are
active in a cooperative fashion and for $D > 1$.
Khodel and Shaginyan\cite{ref6} have recently proposed that a $3D$ system
with long range Coulomb interaction shows a breakdown of FL behavior
with the development of a horizontal inflexion point in the one
particle excitation spectrum at the Fermi momentum ${\bf k}_{_F}$. Such a
feature leads to a restructuring of the Fermi filling and hence of the
Fermi distribution function at $T = 0$ resulting in a finite slope
(rather than a jump) at ${\bf k}_{_F}$.
This phenomenon is very reminiscent of what happens in a BCS superconductor
with a Bogoliubov like excitation spectrum.
It has since been shown\cite{ref7}, that the Khodel-Shaginyan state is an
artifact of the Hartree Fock analysis and that it is indeed unstable
towards a superconducting ground state. Nevertheless it was
recognized\cite{ref7} that such inherent features like potentially
flat portions of the Hartree Fock excitation spectrum should be very
sensitive when including particle -- particle scattering.
Quasi particle broadening should then become a dominant feature and
possibly lead to results which resemble in nothing an ordinary FL.

On the basis of a detailed analyses of the anomalous lattice properties
of high $T_{_C}$ materials we have recently attempted to argue for
the applicability of the Boson-Fermion model to high $T_{C}$
superconductivity\cite{ref8}.
As we shall see in this letter, a detailed analysis of this model shows
features which were recognized as potential ingredients for
the breakdown of FL behavior\cite{ref7} in connection with the study of the
Khodel-Shaginyan model.
The Boson-Fermion model was initially introduced by one of us\cite{ref9} in
order to provide a heuristic description of intermediate coupling
electron-phonon systems and to study the resulting superconducting
properties.
This model was then extensively studied (although in a slightly generalized
form) in connection with high $T_{_C}$ superconductors, in the
superconducting state\cite{ref10}.It is therefore compelling to
examine its normal state properties and to compare them to the
experimentally established anomalous properties -- which are the real
testing grounds for any theory of high $T_{_C}$ superconductivity.

The Boson-Fermion model describes systems with both localized Bosons with
charge $2e$ and itinerant Fermions with charge $e$ on a lattice.
An exchange of a Boson with two Fermions with opposite spins is assumed
and the conservation of charge requires a common chemical potential for
the two species of charge carries. This model exhibits a superconducting
ground state with a BCS like gap in the Fermion excitation
spectrum\cite{ref9} and collective excitations of the Bosons.
What is new in this model as compared to a standard BCS systems is, that well
above $T_{_C}$ a pseudo-gap in the density of states develops which
ultimately is expected to open up into a true gap at the critical
temperature $T_{_C}$. It is this regime of temperature above $T_{_C}$
which is of interest to us here and where clear non-FL
properties manifest themselves.

In this letter we shall study the Boson-Fermion model
\[
H \; = \; (zt - \mu) \sum_{i,\sigma} c_{i\sigma}^{+} c_{i\sigma} \; - \;
t \sum_{<i\neq j>, \sigma} c_{i\sigma}^{+} c_{j\sigma} \; + \; (\Delta_{_B}
- 2 \mu) \sum_{i} b_{i}^{+} b_{i}
\]
\begin{equation}
\; + \; v \sum_{i} [ \; b_{i}^{+} c_{i\downarrow} c_{i\uparrow} \; + \;
c_{i\uparrow}^{+} c_{i \downarrow}^{+} b_{i} \; ]
\label{Equ1}
\end{equation}
for a square lattice.
$t$ denotes the electron hopping integral, $z$ the number of nearest
neighbors, $\Delta_{_B}$ the atomic level of the Bosons (bi-polarons) and
$v$ the strength of the Boson-Fermion exchange coupling.
The commutation relations for the Fermion and Boson operator are
given by
\( \left\{ c_{i\sigma}, c_{j\sigma'} \right\}  =  \delta_{ij}
\delta_{\sigma \sigma'}
\)
and
\(
\left[ b_{i}, b_{j}^{+} \right]  =  \delta_{ij}
\)
respectively.
We evaluate the excitation spectrum of the Fermions in the
normal state in a fully self consistent way\cite{ref11} to second order
in the exchange interaction.
The expressions for the self energies for Fermions and
Bosons are:
\[
\Sigma_{F}({\bf k}, \omega_{n}) \; = \; - \frac{v^{2}}{N} \; \sum_{{\bf q},
\omega_{m}} \; G_{F}(-{\bf k}+{\bf q}, \omega_{m} - \omega_{n}) \;
G_{B}({\bf q}, \omega_{m})
\]
\begin{equation}
\Sigma_{B}({\bf q}, \omega_{m}) \; = \; \frac{v^{2}}{N} \; \sum_{{\bf k},
\omega_{n}} \; G_{F}(-{\bf k}+{\bf q}, -\omega_{n}+\omega_{m}) \; G_{F}(
{\bf k}, \omega_{n})
\label{Equ2}
\end{equation}
This leads to the set of equations for the Fermion and Boson Green's
functions
\[
G_{F}({\bf k}, \omega_{n}) \; = \; \left[ i \omega_{n} \; - \;
\epsilon_{{\bf k}} \; - \; \Sigma_{F}({\bf k}, \omega_{n}) \; \right]^{-1}
\]
\begin{equation}
G_{B}({\bf q}, \omega_{m}) \; = \;
\left[ i \omega_{m} \; - \; E_{0} \; - \;
\Sigma_{B}({\bf q}, \omega_{m})
\; \right]^{-1}
\label{Equ4}
\end{equation}
which are determined selfconsistently, together with the expressions for the
self-energies, Equ.(\ref{Equ2}).
${\bf k}$ and ${\bf q}$ denote the momenta, $\omega_{n}$ and $\omega_{m}$
the Matsubara frequencies for Fermions and Bosons respectively
and N the number of sites.

The unperturbed Fermion dispersion including the chemical potential is
given by $\epsilon_{\bf k} = \xi_{\bf k} -\mu$,
\(
\xi_{\bf k} = t(z - \sum_{\mbox{ {\small \boldmath $\delta$}}}
e^{i {\bf k}
\mbox{ {\small \boldmath $\delta$}}})
\),
with $\delta$ denoting the vectors linking nearest neighbor lattice site.
The unperturbed Boson energies are given by $E_{0} = \Delta_{_B} - 2 \mu$;
the factor two in front of the chemical potential taking account that
each Boson is constituted of two Fermions.

The self-consistent coupled equations (\ref{Equ2}, \ref{Equ4}) are solved
by an iterative procedure in which $G_{F}({\bf k}, \omega_{n})$ and
$G_{B}({\bf q}, \omega_{m})$ are evaluated for a set of Matsubara
frequencies $\omega_{n} = 2 \pi k_{_B} T ( n + {1 \over 2})$ for
$-100 < n < +99$ and $\omega_{m} = 2 \pi k_{_B} T m $ for
$-100 < m < +100$. As usual we compute the difference between the
full and bare Green's functions, so that only a small number of
Matsubara frequencies is necessary. We restrict ourselves in the
present study to summing the ${\bf k}$ and ${\bf q}$ vectors over a
two-dimensional Brillouin zone with a set of $41 \times 41$ equally
spaced vectors for the Bosons as well as the Fermions.
Convergency of the iterative solutions of these self-consistent equations
is obtained relatively fast for temperatures down to $T = 0.0085$ in
units of the Fermionic band width $8t$.
The solutions for the Fermion and Boson Green's functions in terms of the
Matsubara frequencies are then analytically continued to the real
frequencies axis and into the lower half plane using a standard Pad\'e
approximants procedure in order to obtain the poles of
the retarded Green's functions and hence the excitation spectra for
the Fermions and Bosons.

In Fig.(\ref{Fig1})  we plot the Fermion spectral
function
$A_{F}({\bf k}, \omega) = - 2 Im \; G_{F}({\bf k},i \omega_{n} = \omega
+ i0^{+})$
as a function of Fermion momentum ${\bf k}$ which clearly shows how
the spectral weight is split and redistributed in the region near
${\bf k}_{_F}$; ${\bf k}_{_F}$ being defined by
$\Sigma_{F}({\bf k}_{_F}, 0) + \epsilon_{{\bf k}_{_F}} = 0$.
This behavior manifests itself in the Fermionic density of states
$\rho_{F}(\omega)$ in form of the appearance of a pseudo-gap which
deepens with decreasing temperature (Fig.\ref{Fig1}).
Throughout the present work we assume as parameters: $\Delta_{_B} = 0.4$ and
$v = 0.1$ in terms of the bandwidth. $v$ is chosen in such way that
for realistic temperatures (a few hundred degrees Kelvin) we can expect
noticeable effects due to the Boson-Fermion exchange coupling.
The total average  number of particles per site,
\( n_{tot} = n_{_F} + 2n_{_B} = {1 \over N} \sum_{{\bf k}, \sigma}
< c_{{\bf k}, \sigma}^{+} c_{{\bf k}, \sigma} >
+ {2 \over N} \sum_{{\bf q}} < b_{{\bf q}}^{+} b_{{\bf q}} > \),
Bosons and Fermions included,
is taken to be unity.

The spectral functions of the Fermionic single particle excitations near
${\bf k}_{_F}$ have not a simple Lorentzian behavior and are best described by
the continued fraction Pad\'e approximants for $G_{F}({\bf k}, \omega)$.
For practically all ${\bf k}$ vectors around ${\bf k}_{_F}$, one obtains
essentially three poles whose spectral weight adds up to essentially unity.
These three poles are given by a cosin like dispersion -- indicative of the
unrenormalized Fermion excitation spectrum $\epsilon_{{\bf k}}$ -- and
two flat dispersionless branches separated by an energy equivalent to
$\mu$. Solving the Hamiltonian, Equ.(\ref{Equ1}) in the atomic limit
($t \rightarrow 0$ and $\Delta_{_B} < 0$) we find that for states
containing two charge carriers per site there are two distinct eigenstates
having energy
$\epsilon_{\pm} = \Delta_{_B}/2 \pm \sqrt{(\Delta_{_B}/2)^{2} + v^{2}} - 2\mu$,
$\epsilon_{+} \sim 0$
corresponds to a state where the charge carriers exist
predominantely in form of on site bosons -- a bonding state --,
$\epsilon_{-} \sim - 2\mu \sim |\Delta_{_B}|$
corresponds to a state where the charge carriers exist
predominantely in form of two uncorrelated fermions
-- an antibonding state.

The real ($\omega_{{\bf k}}$) and imaginary
($- \gamma_{{\bf k}} /2$) part of
these poles are presented in Fig.(\ref{Fig2}.a, b) and the modulus of
their residues is given in Fig.(\ref{Fig2}.c) as a function of Fermion
momentum. The residues are in general complex numbers whose
imaginary part is  biggest for the modes with largest damping
$\gamma_{{\bf k}}$.
Complex residues in general indicate strong interference between poles,
meaning that the single particle excitations  are no longer described
by a simple pole of the Green's function as in a classical FL.
In the description of deviations from FL properties the frequency dependence
of the imaginary part of the Fermion self-energy,
$\Gamma({\bf k}, \omega) = - 2 Im \Sigma_{F}({\bf k}, i \omega_{n} =
\omega + i0^{+})$, is of capital importance.
For a classical FL we have
$\Gamma({\bf k}, \omega) \sim a T^{2} +
b (\omega - \mu)^{2}$.
In the Boson-Fermion model  the deviations
from such a FL behavior are signifiant as can be seen
from Fig.(\ref{Fig3}) where
we plot $\Gamma({\bf k}_{_F}, \omega)$ for various temperatures.
As in a FL, $\Gamma({\bf k}, \omega)$ turns out to be very weakly dependent
on ${\bf k}$. $\Gamma({\bf k}_{_F}, \omega)$
at finite temperature has a minimum
which occurs for slightly negative frequencies. The reasons for that are
at present not understood.
As the temperature decreases this minimum shifts towards $\omega = 0$
but $\Gamma({\bf k}_{F}, 0) \sim T^{-\alpha}$
increases with decreasing temperature (with $\alpha = 0.086$ for $2D$ and
$\alpha = 0.60$ for $1D$) which is exactly the opposite of what occurs in
a standard FL.
All these manifestly non-FL properties of
the Boson-Fermion model show up in the Fermion distribution function
$n({\bf k})$ only in a relatively minor fashion.
$n({\bf k})$ turns out to be
practically $T$ independent at ${\bf k}_{_F}$, (all lines of
$n({\bf k}, T)$ cross within a very small regime arround ${\bf k}_{_F}$)
but increases substantially with decreasing temperature  for ${\bf k}$
just below ${\bf k}_{_F}$.

The breakdown of FL properties characterized by the disappearance of well
defined quasiparticles near the Fermi energy
is found to go
hand in hand with
the appearance of well defined itinerant excitations of the intrinsically
(bare) localized Bosons
due to a precursor to superfluidity. Such a precursor induced coherence
of intrinsically localized Bosons has recently been studied by
us\cite{ref12} on the basis of the $1D$ Boson-Fermion model.

As the temperature is decreased the low
${\bf q}$ vector  excitations, manifest in the poles of
$G_{B}({\bf q}, \omega)$, acquire the spectrum
$\omega_{{\bf q}}^{B} = \hbar^{2} q^{2} /2 m_{_B}(T)$
with a temperature dependent
mass which decreases upon decreasing the temperature.
We estimate the decrease of $m_{_B}(T)$ to be given by
$m_{_B} = 24.2$, $25.7$, $51.3$ and $195.6$ for
$T = 0.0085$, $0.01$, $0.02$ and $0.05$.

The present analysis of the Boson-Fermion model shows that for $1D$ as well
as for $2D$ the FL properties are totally destroyed.
This is due to strong superconducting fluctuations setting in well above
$T_{_C}$ as can be seen from the pair correlation function -- essentially
determined by the Bose Green's function $G_{B}({\bf q}, \omega)$ --
which has perfect Lorentzian line shape well above $T_{_C}$.
{}From the static part of $G_{B}({\bf q}, \omega)$ we can estimate the
coherence length, determined by
$< b_{{\bf q}}^{+} b_{{\bf q}} > \simeq
a (1 + \xi^{2} q^{2} )^{-1}$
which
varies as $\xi \sim (T - T_{_C})^{-\nu}$ (with $\nu \simeq 1.25$ and
$T_{_C} \simeq 0$ for $2D$ and $\nu \simeq 1.1$ and $T_{_C} \simeq 0$
for $1D$.
The precursor effects of the onset of joint superfluidity of the Bosons
and superconductivity of the Fermions can also be seen in how the
Hugenholtz-Pines theorem ($G_{B}^{-1}({\bf q} = 0, \omega = 0) = 0$ at
$T_{_C}$) is approached {\em i.e.}
$E_{0} + \Sigma_{B}(0, 0) \simeq (T - T_{_C})^{-\alpha}$ with
$\alpha \simeq 2.7$ for $2D$ and $\alpha \simeq 2.4$ for $1D$.
These precursor effects forshadow a Bogoliubov like spectrum
in which flat portions of the Fermionic excitations spectrum appear
near ${\bf k}_{_F}$ which however
are heavily overdamped.
As a consequence, a pseudo-gap in the Fermion density of states begins
to open up well above $T_{_C}$. It gradually deepens and is expected
to become a true gap at $T_{_C}$.
At present our numerical analyses of the normal state
does not permit to prove this but
our previously mean field results\cite{ref13} as well as
preliminary RPA results for the superconducting state suggest
such a behavior.

The fully selfconsistent scheme employed in the present study corresponds
to the simplest approximation in such a treatment. We believe nevertheless
that the results obtained here should be qualitatively correct since vertex
corrections to the selfenergies are of order $v^{4}$ smaller than the
selfenergies taken into account here. Moreover the Hugenholtz-Pines
theorem in the limit $T \rightarrow T_{_C}$ is fulfilled, satisfying
the selfconsistent equation $E_{0} - \Sigma_{B}({\bf q} = 0, \omega = 0) = 0$
for $T = T_{_C} = 0$.

At this points the Boson-Fermion model should be compared with the
negative $U < 0$ Hubbard model for which a two particle resonant state
appears inside the two particle continuum.
Contrary to the Boson-Fermion model, these Bosonic  states are however heavily
overdamped in the $U < 0$ Hubbard model\cite{ref14}
 and thus are not expected  to control the superconducting state via their
Bose Einstein condensation.
The absence of well defined bosonic excitations
does not cause single Fermion excitations to become overdamped
and the $U < 0$ Hubbard model, in spite of many similarities with
the Boson-Fermion model thus remains a standard
FL in the intermediate coupling regime {\em i.e.} $U \simeq 4t$\cite{ref14}.

Concerning the applicability of the Boson-Fermion model to high $T_{_C}$
superconductors, it shows at least one predominant feature
which a series of experiments
suggest; namely the existence of a pseudo gap and non-FL behavior
in the normal state.
NMR, Knight shift, susceptibility measurements, optical conductivity,
specific heat,
etc\cite{ref15} all suggest such a pseudo-gap in the underdoped samples.
The behavior of the width of the fermionic spectral function,
measured by photo-emission,
favors however more the picture of a marginal
FL than a completely destroyed one as discussed in this letter.
In this connection we should however remember that the Boson-Fermion model
examined here is the prototype of this model which does not take into
account any details of the cristallin lattice and electronic structure
of real materials.
Further work on this matter, also including shortrange spin correlations
in the Fermionic subsystem, will eventually elucidate
these questions.

\acknowledgements

We thank K. Matho and T. Kostyrko for valuable conversations.

\begin{figure}
\caption{Density of states and the spectral functions of the electrons
for a set of $k$ vectors,
$k = k_{x} = k_{y} = \frac{2\pi}{41} \times [0... 11]$. The results for
additional equidistant ${\bf k}$ vectors was obtained by an interpolation
of the selfenergy.
In the inset we show the deepening of the pseudo-gap,
centered at energy $0$, as the temperature
decreases, for three characteristic temperatures.}
\label{Fig1}
\end{figure}


\begin{figure}
\caption{a) The real part of the poles of the Fermion Green's function
$G_{F}({\bf k}, \omega)$ for ${\bf k}$ vectors along the diagonal of
the Brilloin zone $k = k_{x} = k_{y} = \frac{2 \pi}{41} \times [0... 20]$.
b) The imaginary part divided by
the real part of the poles of $G_{F}({\bf k}, \omega)$.
c) The modulus of the residues of the poles of $G_{F}({\bf k}, \omega)$ }
\label{Fig2}
\end{figure}

\begin{figure}
\caption{The imaginary part of the electron self-energy for $k_{x} = k_{y}
= k_{_F}$ for various temperatures. $\Gamma({\bf k}_{_F}, 0)$ increases with
decreasing temperature.}
\label{Fig3}
\end{figure}

\end{document}